\documentclass{phbauth}
\usepackage{graphicx,epsfig,float}

\begin{document}

\begin{frontmatter}

\title{Fermi surface of Sr$_2$RuO$_4$ from angle resolved photoemission}

\author{A. Damascelli\thanksref{thank1}$^{a,\!}$},
\author{K.M. Shen$^a$, D.H. Lu$^a$, N.P. Armitage$^a$, F. Ronning$^a$, D.L. Feng$^a$,}
\author{C. Kim$^a$, Z.-X. Shen$^a$, T. Kimura$^b$, Y. Tokura$^b$, Z.Q. Mao$^c$, and Y. Maeno$^c$}

\address{$^a$Dept. of Physics, Applied Physics, and Stanford Synchrotron
        Radiation Lab., Stanford University, Stanford, CA 94305}
\address{$^b$Dept. of Applied Physics, The
 University of Tokyo, Tokyo 113-8656, and JRCAT, Tsukuba, 305-0046, Japan}
\address{$^c$Dept. of Physics, Kyoto University, Kyoto 606-8502, and CREST-JST, Kawagushi,
Saitama 332-0012, Japan}

\thanks[thank1]{Corresponding author. {\it E-mail}: damascel@stanford.edu}

\begin{abstract}
We investigated the low-energy electronic structure of
Sr$_2$RuO$_4$ by angle-resolved photoemission spectroscopy (ARPES)
focusing, in particular, on the controversial issues of the Fermi
surface (FS) topology and of the extended van Hove singularity
(evHs). A detailed study of temperature and photon energy
dependence demonstrates the surface-state nature of the electronic
band responsible for the evHs, and may suggest the presence of
ferromagnetic correlations at the surface. In addition, we
detected a replica of the primary FS that indicates a
$\sqrt{2}\!\times\!\sqrt{2}$ surface reconstruction, as confirmed
by LEED. In light of these findings, the FS determined by ARPES is
consistent with LDA and de Haas-van Alphen (dHvA) results, and it
provides additional information on the detailed shape of the
$\alpha$, $\beta$ and $\gamma$ sheets. Also, at this stage,
because of the strong sensitivity of the cleaved surface, we could
not find any robust signature of the crossover from 2D to 3D Fermi
liquid behavior in our ARPES measurements for $T\!<\!130$ K.
\end{abstract}

\end{frontmatter}

The determination of the Fermi surface topology of Sr$_2$RuO$_4$,
by ARPES has been matter of intense debate. The resolution of this
controversy is important not only for the physics of Sr$_2$RuO$_4$
{\it per se}, but also as a reliability test for FSs determined by
ARPES, especially on those correlated systems where photoemission
is the only available probe. In this context, Sr$_2$RuO$_4$ is a
particularly interesting material because it can also be
investigated with dHvA experiments, contrary to the cuprates, thus
providing a direct test for the ARPES results. Whereas dHvA
analysis \cite{mackenzie}, in agreement with LDA band-structure
calculations \cite{oguchi,singh}, indicates two electron pockets
$\beta$, and $\gamma$ centered at the $\Gamma$ point, and a hole
pocket $\alpha$ at the X point [as sketched in 1/4 of the
projected zone (PZ) in Fig.1], early ARPES measurements suggested
a different picture: one electronlike FS ($\beta$) at the $\Gamma$
point and two hole pockets ($\gamma$, and $\alpha$) at the X point
\cite{lu,yokoya}. The difference comes from the detection by ARPES
of an intense, weakly dispersive feature at the M point just below
$E_F$, that was interpreted as an evHs singularity pushed down
below $E_F$ by electron-electron correlations \cite{lu,yokoya}.
The existence of the evHs was questioned in a later photoemission
paper \cite{puchkov}, in which the feature detected at the M point
was suggested to be a surface state (SS). Very recently
\cite{boer}, it has been proposed that ARPES could be probing
ferromagnetic correlations of the surface, which would result in
two $\gamma$-FS with different topology (hole and electronlike,
respectively, for majority and minority spin direction).
\begin{figure}[t!]
\centerline{\epsfig{figure=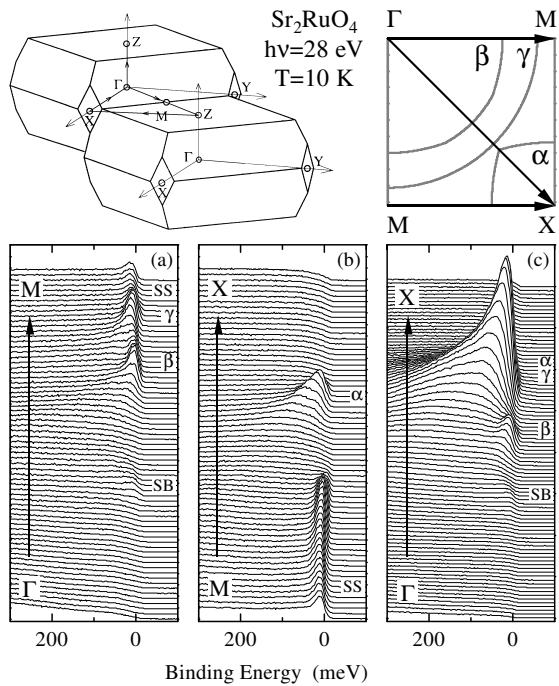,width=0.98\linewidth,clip=}}
\vspace{-.2cm} \caption{ARPES spectra from Sr$_2$RuO$_4$ along the
symmetry lines $\Gamma$-M, M-X, and $\Gamma$-X. As shown in the
sketch of the 3D Brillouin zone (BZ), M is the midpoint along
$\Gamma$-Z and, together with $\Gamma$ and X, defines the 2D
projected zone (PZ). In the quadrant of the 2D PZ, $\alpha$,
$\beta$ and $\gamma$ sheets of FS are indicated together with the
experimentally measured cuts.} \label{edc}
\end{figure}

We investigated in detail the electronic structure of
Sr$_2$RuO$_4$. By varying the incident photon energy and the
temperature at which the samples were cleaved, we confirm the SS
nature of the near-$E_F$ peak detected at the M point, and we
identify an additional dispersive feature related to the `missing'
electronlike FS ($\gamma$). Full understanding of the data can be
achieved only by recognizing the presence of shadow bands (SB),
due to a $\sqrt{2}\times\!\sqrt{2}$ surface reconstruction (as
confirmed by LEED). The FS thus determined by ARPES is consistent
with the dHvA results \cite{mackenzie}. A detailed temperature
dependent analysis is complicated by the strong sensitivity of the
surface: the latter rapidly degrades upon increasing temperature.
At this stage, we could not find in our ARPES spectra any reliable
signature of the crossover from 2D to 3D Fermi liquid behavior,
reported for $T\!<\!130$ K \cite{katsufuji,maeno}.

High energy (14 meV) and angular
($0.5^{\circ}\!\times\!0.3^{\circ}$) resolution ARPES data was
taken at SSRL, on the normal incidence monochromator beam line
equipped with a SES-200 electron analyzer. Sr$_2$RuO$_4$ single
crystals were oriented by conventional Laue diffraction, and then
cleaved {\it in situ} with a base pressure better than
5x10$^{-11}$ torr. More details on the experiment and on the
normalization methods can be found in Ref. \cite{damascelli}.

Fig.1 presents, for Sr$_2$RuO$_4$ cleaved and measured at 10 K,
energy distribution curves (EDCs) along the high-symmetry
directions (all detected features are labeled following the
assignment given in the paper). Along $\Gamma$-M, two peaks emerge
from the background, disperse towards $E_F$, and cross it before
the M point, defining $\beta$ and $\gamma$ electronlike pockets
(Fig.1a). Along M-X, a peak approaches and crosses $E_F$ before
the X point defining, in this case, the hole pocket $\alpha$
centered at X (Fig.1b). Similar results were obtained along
$\Gamma$-X (Fig.1c): the $\beta$ pocket is clearly resolved, while
$\alpha$ and $\gamma$ crossings are almost coincident. In
addition, we identify a weak feature that shows dispersion
opposite to the primary peaks along $\Gamma$-M and $\Gamma$-X (SB,
see below). Around the M point, we can observe the sharp peak (SS)
that was initially associated to a holelike sheet of FS centered
at X \cite{lu,yokoya}.

In the following discussion, we will concentrate on the features
observed near the M point, which are relevant to the controversy
concerning the character of the $\gamma$ sheet of FS. In
particular, we will show that working with sufficient momentum
resolution, both the $\beta$ and $\gamma$ electronlike pockets
(predicted by LDA calculations \cite{oguchi,singh}) are clearly
resolved in the ARPES spectra. In order to address this issue, we
measured the M point region (with cuts along $\Gamma$-M-$\Gamma$)
varying the incident photon energy between 16 and 29 eV, in steps
of 1 eV. Here, we covered the location of $\beta$ and $\gamma$
pockets in both first and second zone (i.e., four $E_F$ crossings
ought to be observed). From the EDCs shown in Fig.2 for 16, 22,
and 28 eV, we can see that the cross sections of SS, $\beta$ and
{\it in particular} $\gamma$ exhibit a strong (and different)
dependence on photon energy. At 28 eV, $\beta$ and $\gamma$
crossings can be individually identified in the EDCs. Owing to the
high momentum resolution we can now follow the dispersion of the
$\gamma$ peaks until the leading edge midpoints are located beyond
$E_F$. After that, the peaks lose weight and disappear, defining
the $k_F$ vectors for the electronlike $\gamma$ pockets. Right at
$k_F$ we can resolve a double structure which then reduces to the
non dispersive feature (SS) located 11 meV below $E_F$. The
difference between the 28 eV results and those obtained at 16 or
22 eV is striking. At low photon energies, the $\beta$ crossings
are still clearly visible. On the other hand, we can follow only
the initial dispersion of the $\gamma$ peaks, before they merge
with the SS feature, giving the impression of an evHs. At 16 eV it
is impossible to identify the $\gamma$ crossings. At 22 eV the
location of the leading edge midpoints is at best suggestive of
the presence of the $\gamma$ crossings.
\begin{figure}[t!]
\centerline{\epsfig{figure=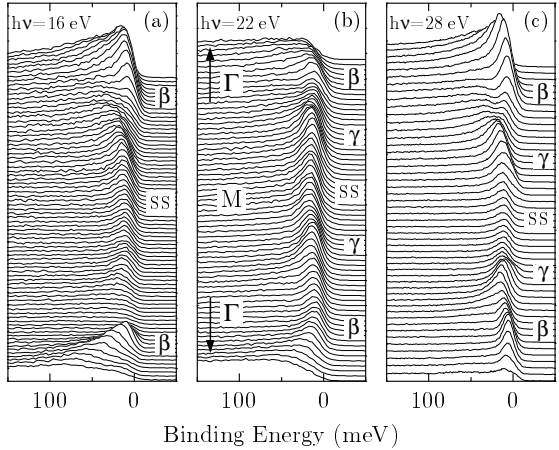,width=0.98\linewidth,clip=}}
\vspace{-.2cm} \caption{ARPES spectra along $\Gamma$-M-$\Gamma$,
at different photon energies. The cuts are centered at the M point
and extend beyond the $\gamma$ and $\beta$ FS in both first and
second zone.} \label{hv}
\end{figure}

\begin{figure}[b!]
\centerline{\epsfig{figure=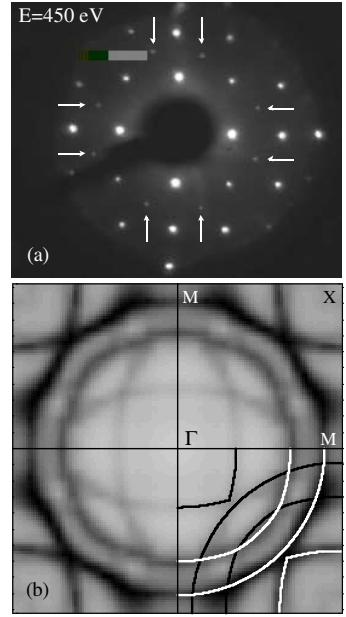,width=0.6\linewidth,clip=}}
\vspace{-.2cm} \caption{LEED pattern (a) measured at 10 K with 450
eV electrons. The arrows indicate superlattice reflections due to
$\sqrt2\!\times\!\sqrt2$ surface reconstruction. Fermi energy
intensity map (b), obtained by integrating the EDCs over an energy
window of $\pm10$ meV about the chemical potential. Primary
$\alpha$, $\beta$ and $\gamma$ sheets of FS are marked by white
lines, and replica due to surface reconstruction by black lines.
Data was taken on Sr$_2$RuO$_4$ cleaved at 10 K.} \label{leed}
\end{figure}

In order to have a full picture of the relevant issues to be
addressed, let us proceed to the discussion of the FS mapping
obtained by integrating the EDCs over a narrow energy window about
$E_F$. As the electron spectral function, multiplied by the Fermi
function, reaches its maximum at $E_F$ when a band crosses the
Fermi energy, the FS is identified by the local maxima of the
intensity map. Fig.3b shows the $E_F$ intensity map obtained at 28
eV on a Sr$_2$RuO$_4$ single crystal cleaved and measured at 10 K.
The actual EDCs were taken over more than a full quadrant of the
PZ with a resolution of 0.3$^{\circ}$ (1$^{\circ}$) in the
horizontal (vertical) direction. EDCs were then integrated over an
energy window of $\pm10$ meV about the chemical potential. The
resulting map was then symmetrized with respect to the diagonal
$\Gamma$-X (to compensate for the different resolutions along
horizontal and vertical directions). The $\alpha$, $\beta$, and
$\gamma$ sheets of FS are clearly resolved, and are marked by
white lines in Fig.3b. In addition, Fig.3b shows some unexpected
features: besides the diffuse intensity around the M point due to
the presence of the SS band, there are weak but yet well defined
profiles (marked in black). They can be recognized as a replica of
the primary FS, and are related to the weak SB features detected
in the EDCs along the high-symmetry lines (Fig.1a and 1c). This
result is reminiscent of the situation found in
Bi$_2$Sr$_2$CaCu$_2$O$_8$ where similar shadow bands are possibly
related to AF correlations, or to the presence of two formula
units in the unit cell. On the other hand, in Sr$_2$RuO$_4$ the
origin of the SB is completely different: inspection with LEED
reveals superlattice reflections corresponding to a
$\sqrt{2}\!\times\!\sqrt{2}$ surface reconstruction (see Fig.3a),
which is responsible for the folding of the primary electronic
structure with respect to the M-M direction. Quantitative LEED
analysis of the surface structure shows a 9$^\circ$ rotation of
the RuO$_6$ octahedra around the surface normal, which leads to
the enlargement of the in-plane unit-cell dimensions by
$\sqrt2\!\times\!\sqrt2$ over that of the bulk \cite{leed}. This
reconstruction, which is absent in the cuprates, reveals an
intrinsic instability of the cleaved surface of Sr$_2$RuO$_4$, and
should be taken into account as the origin of possible artifacts
also in other surface sensitive measurements like, e.g., scanning
tunneling microscopy.
\begin{figure*}[t!]
\centerline{\epsfig{figure=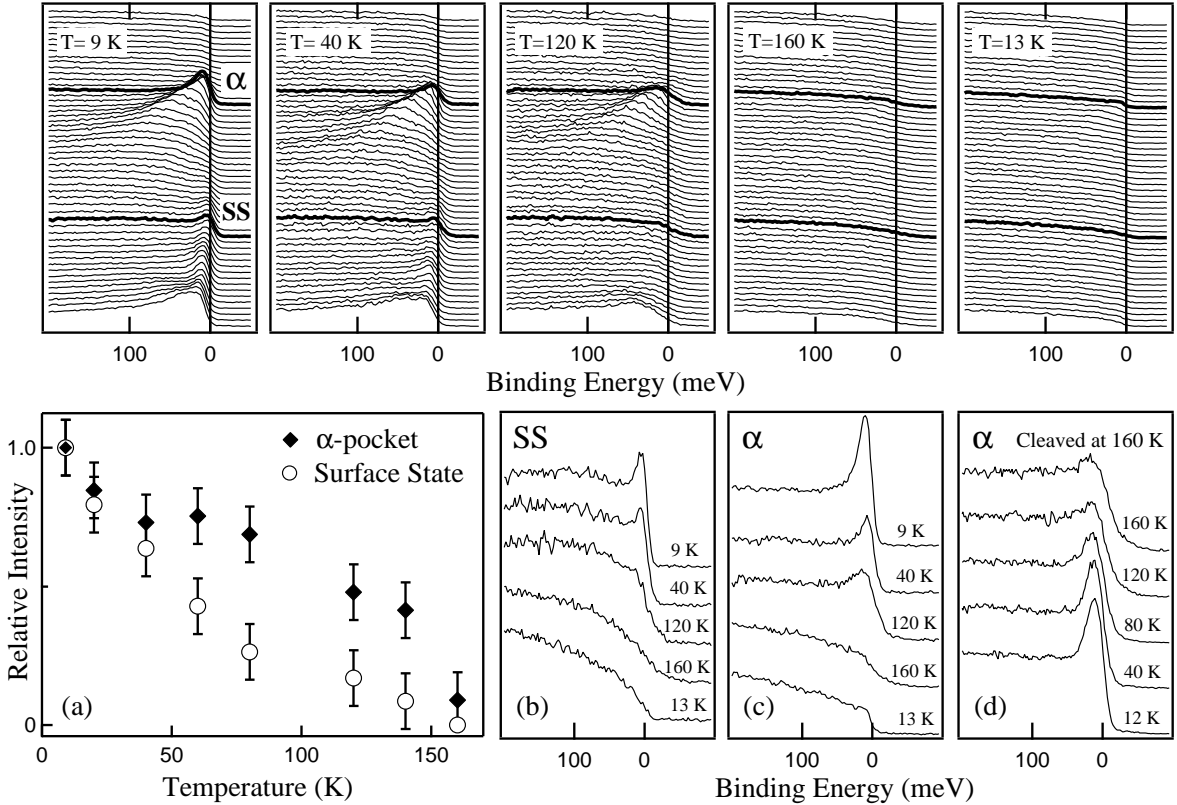,width=0.98\linewidth,clip=}}
\vspace{-.2cm} \caption{Top panels: EDCs along M-X measured upon
increasing the temperature from 9 to 160 K. The absence of
features in the EDCs taken at 13 K, at the end of the run,
indicates the complete degradation of the surface. An enlarged
view of the T-dependence for SS and $\alpha$ pocket (thick lines
in the top panels) is given in (b) and (c), respectively. These
results, as emphasize by the normalized relative peak intensity of
SS and $\alpha$ pocket plotted versus temperature (a), show the
higher sensitivity of SS on surface degradation. Panel (d): EDCs
for the $\alpha$ pocket taken, upon reducing T, on a sample
cleaved at 160 K. Note that a peak is present already at 160 K,
i.e., well above the 2D-3D crossover temperature (130 K).}
\label{leed}
\end{figure*}

By inspecting the M point (Fig.3b), it becomes now clear why the
investigation of this k-space region with ARPES has been so
controversial: in addition to the weakly dispersive SS feature
(Fig.1 and 2), there are several sheets of FS (primary and
`folded'). At this point, the obvious question is: what is the
exact nature of the SS feature? In order to verify that this
feature indeed arises from a surface state \cite{puchkov}, we
investigated its sensitivity to surface degradation by cycling the
temperature between 10 and 200 K.
\begin{figure*}[t!]
\centerline{\epsfig{figure=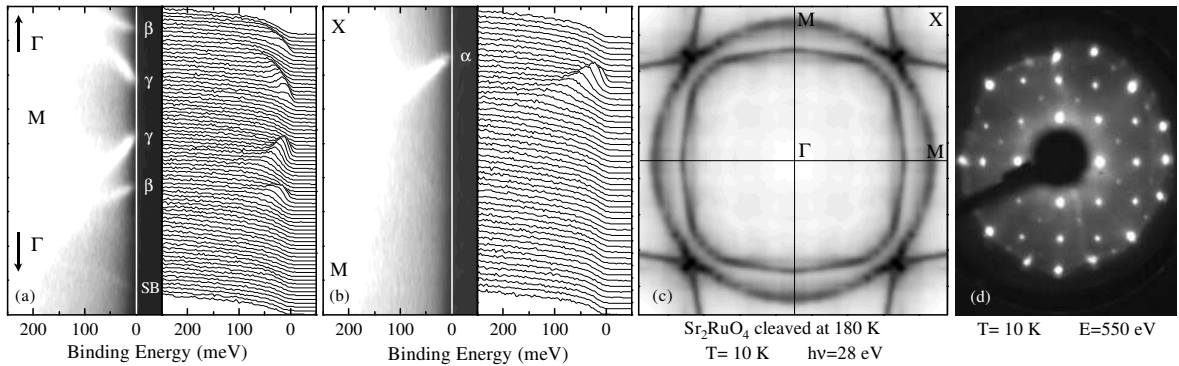,width=0.98\linewidth,clip=}}
\vspace{-.2cm} \caption{EDCs and intensity plot
$I(\mathbf{k},\omega)$ along the high-symmetry directions
$\Gamma$-M-$\Gamma$, and M-X (panel a and b, respectively). Panel
c: $E_F$ intensity map obtained by integrating the EDCs over an
energy window of $\pm10$ meV about the chemical potential. Panel
d: LEED pattern recorded at the end of the FS mapping. All data
was taken at 10 K on a Sr$_2$RuO$_4$ single crystal cleaved {\it
in situ} at 180 K. Note that whereas the surface reconstruction
(i.e., superlattice reflections in the LEED pattern, panel d), and
the shadow bands produced by the consequent folding of the primary
electronic structure (panel a) are still observable, the surface
state peak at the M point has completely disappeared (panel a and
b).} \label{nss}
\end{figure*}

In Fig. 4 (top panels), where EDCs measured along M-X are
presented, we can observe both $\alpha$-pocket and SS peaks. Upon
increasing the temperature from 9 to 160 K, both peaks show a
large and monotonic decrease of spectral weight and eventually
disappear  [an enlarged view of the EDCs for SS and $\alpha$
pocket, at the positions marked by thick lines in the top panels,
is given in Fig. 4b and 4c, respectively]. At this point one may
wonder whether this is an intrinsic effect reflecting the 2D-3D
crossover reported for $T\!\simeq\!130$ K \cite{katsufuji,maeno}.
In that case the intensity of the peaks should once again increase
upon reducing the temperature. Unfortunately this check shows that
the observed effect is irreversible and only reflects the
reactivity of the surface of Sr$_2$RuO$_4$: EDCs re-taken at 13 K
are completely featureless, indicating the complete degradation of
the surface (see Fig. 4, top right panel). Furthermore these
results, as emphasized by the relative peak intensity of SS and
$\alpha$ pocket plotted versus temperature in Fig. 4a (data
normalized at 9 K), show the higher sensitivity of the SS to
surface degradation and, therefore, demonstrate the surface state
nature of this feature. In order to gain more insights on the
issue of the 2D-3D crossover, we repeated the experiment upon
decreasing the temperature, on a sample cleaved at 160 K (Fig. 4d
presents EDCs relative to the $\alpha$-pocket). The data shows a
considerable temperature dependence which, however, seems to be
mostly due to the thermal broadening of the Fermi function,
although additional effects can not be fully excluded at this
stage. As a peak is clearly present already at 160 K (i.e., well
above the 2D-3D crossover temperature of 130 K), it seems a fair
statement to say that no strong effect due to the crossover is
detected in the ARPES spectra.

Because the SS peak is suppressed much faster than all other
features, by cleaving the crystals at 180 K and immediately
cooling to 10 K, we suppressed the SS, while only weakly affecting
the intensity of the other electronic states. A more sizable
effect is observed on the SB, confirming a certain degree of
surface degradation. However, the latter was not too severe, as
demonstrated by the LEED pattern taken after the measurements
which still clearly shows the surface reconstruction (Fig.5d).
EDCs measured near the M-point at 10 K (on a sample cleaved at 180
K), and corresponding intensity plots $I(k,\omega)$ are shown in
Fig.5a and 5b. No clear signature of the SS is detected, and the
identification of the Fermi vectors of $\alpha$, $\beta$, and
$\gamma$ pockets is now straightforward.

Performing a complete mapping on a sample cleaved at 180 K, we
obtained an extremely well defined FS (Fig.5c). With the surface
slightly degraded, we expect to see less of the relative intensity
coming from SB and SS. At the same time, we might have expected
the primary FS to be less well defined, which is precisely
opposite to what is observed. The FS shown in Fig.5c is in very
good agreement with LDA calculations \cite{oguchi,singh} and dHvA
experiments \cite{mackenzie}. The number of electrons contained in
the FS adds up to a total of 4 (within an accuracy of 1\%), in
accordance with the Luttinger theorem (for the FS determined on
samples cleaved at 10 K the accuracy in the electron counting
reduces to 3\% due to the additional intensity of folded bands and
surface state).

Our results confirm the surface state nature of the SS peak
detected at the M point. The comparison of Figs.3b and 5c suggests
that a surface contribution to the total intensity is responsible
also for the less well defined FS observed on samples cleaved at
10 K. At this point, one might speculate that these findings are a
signature of the surface ferromagnetism (FM) recently proposed for
Sr$_2$RuO$_4$ \cite{boer}. In this case, two different FSs should
be expected for minority and majority spin directions \cite{boer},
resulting in: (i) additional $E_F$-weight near M due to the
presence of a holelike $\gamma$ pocket for the majority spin; (ii)
overall momentum broadening of the FS contours because the
$\alpha$, $\beta$ and $\gamma$ sheets for the two spin populations
are slightly displaced from each other in the rest of the BZ. In
this scenario, a slight degradation of the surface might suppress
the signal related to FM correlations, due to the introduced
disorder. The resulting FS would then be representative of the
non-magnetic electronic structure of the bulk (Fig.5c). The
hypothesis of a FM surface seems plausible because the instability
of a non magnetic surface against FM order is not only indicated
by {\it ab initio} calculations \cite{boer}, but it may also be
related to the lattice instability evidenced by the surface
reconstruction.

In summary, our investigation provides direct evidence for the
surface state nature of the weakly dispersive feature detected at
the M point (a possible fingerprint of a FM surface). From both
ARPES and LEED, we found that a $\sqrt{2}\!\times\!\sqrt{2}$
surface reconstruction occurs in cleaved Sr$_2$RuO$_4$, resulting
in the folding of the primary electronic structure. Taking these
findings into account, the FS determined by ARPES is consistent
with dHvA results, providing additional information on the
detailed shape of $\alpha$, $\beta$ and $\gamma$ pockets. No clear
evidence was found in favor of a crossover from 2D to 3D Fermi
liquid behavior for $T\!<\!130$.

We gratefully acknowledge C. Bergemann, M. Braden, Ismail, and T.
Mizokawa for useful discussions. SSRL is operated by the DOE
office of Basic Energy Research, Division of Chemical Sciences.
The office's division of Material Science provided support for
this research. The Stanford work is also supported by NSF grant
DMR9705210 and ONR grant N00014-98-1-0195.

\end{document}